# Machine-Learning-Assisted and Real-Time-Feedback-Controlled Growth of InAs/GaAs Quantum Dots

Chao Shen,[†,§] Wenkang Zhan,[†,‡] Kaiyao Xin,[‡,#] Manyang Li,[†,‡] Zhenyu Sun,[†,‡] Hui Cong,[‡,ⵏ] Chi Xu,[‡,ⵏ] Jian Tang,[ⵏ] Zhaofeng Wu,[§] Bo Xu,[†,‡] Zhongming Wei,[‡,#] Chunlai Xue,[‡,ⵏ] Chao Zhao,*[,†,‡] and Zhanguo Wang[†,‡]

[†] Key Laboratory of Semiconductor Materials Science, Institute of Semiconductors, Chinese Academy of Sciences & Beijing Key Laboratory of Low Dimensional Semiconductor Materials and Devices, Beijing 100083, China

[‡] College of Materials Science and Opto-Electronic Technology, University of Chinese Academy of Science, Beijing 101804, China

[#] State Key Laboratory of Superlattices and Microstructures, Institute of Semiconductors, Chinese Academy of Sciences, Beijing 100083, China

[ⵏ] State Key Laboratory on Integrated Optoelectronics, Institute of Semiconductors, Chinese Academy of Sciences, Beijing 100083, China

[§] School of Physics Science and Technology, Xinjiang University, Urumqi, Xinjiang 830046, China

‖ School of Physical and Electronic Engineering, Yancheng Teachers University, Yancheng 224002, China

*Email: zhaochao@semi.ac.cn

# Abstract

Self-assembled InAs/GaAs quantum dots (QDs) have properties highly valuable for developing various optoelectronic devices such as QD lasers and single photon sources. The applications strongly rely on the density and quality of these dots, which has motivated studies of the growth process control to realize high-quality epi-wafers and devices. Establishing the process parameters in molecular beam epitaxy (MBE) for a specific density of QDs is a multidimensional optimization challenge, usually addressed through time-consuming and iterative trial-and-error. Here, we report a real-time feedback control method to realize the growth of QDs with arbitrary density, which is fully automated and intelligent. We developed a machine learning (ML) model named 3D ResNet 50 trained using reflection high-energy electron diffraction (RHEED) videos as input instead of static images and providing real-time feedback on surface morphologies for process control. As a result, we demonstrated that ML from previous growth could predict the post-growth density of QDs, by successfully tuning the QD densities in near-real time from $1.5 \times 10^{10}$ cm$^{-2}$ down to $3.8 \times 10^{8}$ cm$^{-2}$ or up to $1.4 \times 10^{11}$ cm$^{-2}$. Compared to traditional methods, our approach, with in situ tuning capabilities and excellent reliability, can dramatically expedite the material optimization process and improve the reproducibility of MBE, constituting significant progress for thin film growth techniques. The concepts and methodologies proved feasible in this work are promising to be applied to a variety of material growth processes, which will revolutionize semiconductor manufacturing for optoelectronic and microelectronic industries.

## Introduction

Self-assembled quantum dots (QDs) have attracted great interest due to their applications in various optoelectronic devices.[1] The so-called Stranski-Krastanow (SK) mode in molecular beam epitaxy (MBE) is widely used for growing these high-quality QDs.[2] For specific applications, such as QD lasers, high QD densities are required; while low-density QDs are necessary for other applications, such as single photon sources.[1] However, the outcomes of any QD growth process are a complex function of a large number of variables including the substrate temperature, III/V ratio, and growth rate, etc. Building a comprehensive analytical model that describes complex physical processes occurring during growth is an intractable problem.[3] The optimization of material growth largely depends on the skills and experience of MBE researchers. Time-consuming trial-and-error testing is inevitably required to establish optimal process parameters for the intended material specification. It is needed to build a real-time connection between in situ characterization and material growth status.

Machine learning (ML) is revolutionary due to its exceptional capability for pattern recognition and its potential in approximating the empirical functions of complex systems. It enables researchers to extract valuable insights and identify hidden patterns from large datasets, leading to a better understanding of complex phenomena required to build predictive models, generate new hypotheses, and determine optimal growth conditions for MBE.[4, 5, 6, 7, 8] It offers an alternative approach where the growth outcomes for an arbitrary set of parameters can be accurately predicted via a trained neural network, which has been applied to extract film thickness and growth rate information.[9, 10] Moreover, by enabling the direct adjustment of parameters during material growth, ML-based in situ control can detect and correct any deviation from expected values in a timely manner.[11, 12, 13] Meng et al. have utilized a feedback control to adjust cell temperatures after an amount of film being deposited, thereby capable of regulating $Al_xGa_{1-x}As$ compositions in situ.[13] However, these remain posteriori ML-based approaches, since they require the completion of the growth.[10] Therefore, once the sample characterization results deviate from expectation, it becomes challenging and time-consuming to identify reasons.[11]

To fully exploit the effectiveness of real-time connection based on ML, an in-situ characterization with continuous and transient working mode is highly desirable to unveil the material status. Reflection high-energy electron diffraction (RHEED) has been widely used to capture a wealth of growth information in situ.[14] However, it still faces the challenges of extracting information from noisy and overlapping images. Traditionally, identifying RHEED patterns during material growth was mainly depended on the experience of growers. Progress has been made in the automatic classification of RHEED patterns through ML that incorporates principal component analysis (PCA) and clustering algorithms.[9, 14, 15, 16, 17] It surpasses human analysis when a single static RHEED image is gathered with the substrate held at a fixed angle.[16, 18] However, this may result in an incomplete utilization of the temporal information available in RHEED videos taken with the substrate rotating continuously.[16, 19, 20] Although few researchers have highlighted the importance of analyzing RHEED videos, their ML models were still based on PCA and encoder-decoder architecture with a single image input.[16, 21] To date, it is still hard to build a real-time connection between in situ characterization and material growth status, which is a prospective way to realize the precise control on material specifications.

In this work, we proposed a real-time feedback control based on ML to connect in situ RHEED videos and QD density, automatically and intelligently achieving a low-cost, efficient, reliable, and reproduceable growth of QDs with required density. By investigating the temporal evolution of RHEED information during the growth of InAs QDs on GaAs substrates, the real-time status of the sample was better reflected. We applied two structurally identical 3D ResNet 50 models in ML deployment, which takes three-dimensional variables as inputs to determine the QD formation and classify density.[22, 23] We showed that ML can simulate and predict the post-growth material specifications, and based on which, growth parameters are continuously monitored and adjusted. As a result, the QD densities were successfully tuned from $1.5 \times 10^{10}$ cm$^{-2}$ down to $3.8 \times 10^{8}$ cm$^{-2}$, and up to $1.4 \times 10^{11}$ cm$^{-2}$, respectively. The methodology can also be adapted to other QD systems, such as the droplet QDs, the local droplet etching, and manhole infilling QDs with distinct RHEED features.[24, 25, 26, 27] The effectiveness of our approach by taking full advantages of in situ characterization marks a significant achievement of establishing a precise growth control scheme, with the capabilities and potentials of being extended to large-scale material growth, reducing the impact on material growth due to instability and uncertainty of MBE operation, shortening the parameter optimization cycle, and improving the final yield of material growth.

## Results

**Sample structure and data labeling**

Our method relied on an existing database of growth parameters with corresponding QD characteristics, which helped us decide the growth parameters needed to achieve a target density for QDs. The sample structure is shown in Figure 1a, with density results of 30 QD samples summarized in Figure 1b, which are correlated with the RHEED videos. Each QD growth was repeated at least 3 times, leading to the generation of 120 RHEED videos. The samples grown without QDs have flat surfaces determined by an atomic force microscope (AFM) as shown in Figures 1c, corresponding to streaky RHEED patterns in Figure 1g, which was artificially labeled with "zero" referring to the density of QDs on the surface.[27] With the increase of QD density, it was found that the RHEED gradually converted from streaky to spotty patterns. The QDs density of approximately $2.9 \times 10^9$ cm$^{-2}$ from Figure 1d corresponds to the RHEED pattern consisted of streaks and spots in Figure 1h.[27] As shown in Figure 1e and 1i, with the further increase of QDs density reaching $1.9 \times 10^{10}$ cm$^{-2}$, the arrangement of RHEED pattern showed typical spot features.[28, 29, 30] As the density exceeds $4 \times 10^{10}$ cm$^{-2}$, the spot features become more rounded and show higher brightness. Upon conducting a more thorough analysis of the RHEED patterns, we identified the patterns transition from having both streaks and spots to only spots, corresponding to a density of $1 \times 10^{10}$ cm$^{-2}$ in AFM images. So, we assigned the "low" label to densities below $1 \times 10^{10}$ cm$^{-2}$, the "middle" label to densities ranging from $1 \times 10^{10}$ cm$^{-2}$ to $4 \times 10^{10}$ cm$^{-2}$ and labeled densities exceeding $4 \times 10^{10}$ cm$^{-2}$ as "high". (see Supplementary Information for RHEED characteristics of QDs with different labels, S1). Figures 1f and 1j show the typical AFM and RHEED images after the QD formation, respectively, with the QD density of approximately $1.0 \times 10^{11}$ cm$^{-2}$. It is worth noting that we keep the labels to the RHEED videos the same before and after the QD formation. The RHEED information before the QD formation can be used to determine the resulting QD density grown under the same conditions, enabling us to adjust the material growth parameters before the QD formation.

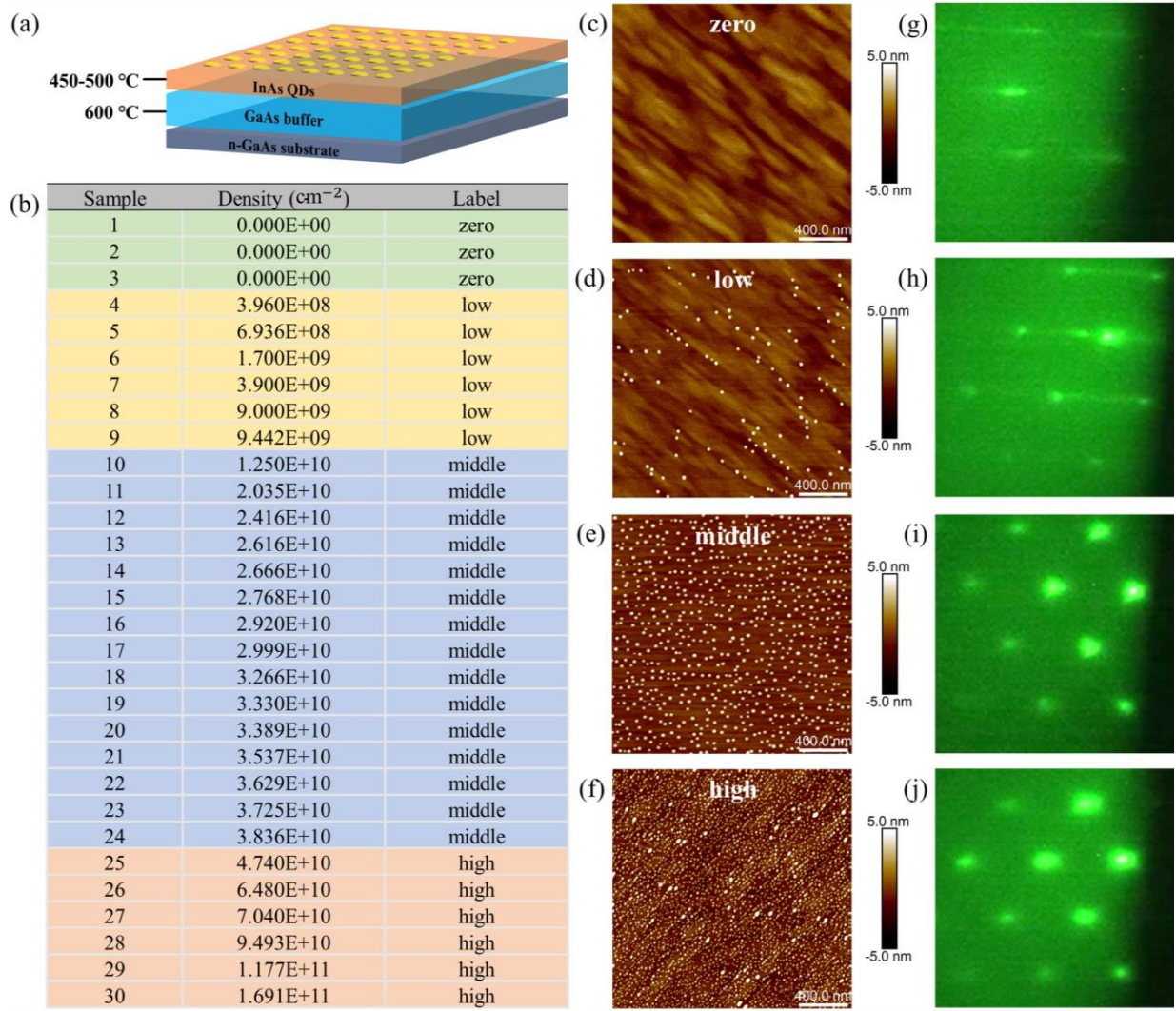

**Fig. 1: QD sample structure and data labeling.** (a) The schematic of the sample structure. (b) QD densities and labels assigned for the samples in the dataset, the density of QDs was determined by counting the number of QDs in the AFM image. (c-f) The 2 μm × 2 μm AFM images of QDs with varied labels. (g-j) Corresponding RHEED images of QDs with varied labels.

**Program framework and video processing**

The design framework of our program is illustrated in Figure 2a. An appropriate scheme was designed to pre-process the RHEED video data. Then, a ML model was selected based on the pre-processing results. In addition, we also determined the way to adjust parameters according to output results of the model.

RHEED videos were first deconstructed into multiple temporal images and utilized as input for our model, which is a technique for breaking down video processing, offering a more versatile and efficient approach to analyze and process information in dynamic scenes.[31] The original image collected has uncompressed 4 channels of 8-bit depth color with a resolution of 1920 × 1200. The software cropped a square area from the image, as shown in Figure 2b, immediately compressed to 300 × 300 by zero-order sampling and converted it to a single 4 × 300 × 300 matrix (see Supplementary Information for the cropping area of RHEED images, S2). To efficiently utilize the temporal information within RHEED data, we used the latest RHEED image as a starting point, acquired an additional seven consecutive images before, and bundled as the original RHEED dataset for the model, as shown in Figure 2c.[32] Subsequently, each image was converted to a single-channel 8-bit grayscale image based on the luminance information,[16, 21] and stitched into an 8 × 300 × 300 3D matrix through shift registers, as shown in Figure 2d. Additionally, we modified the convolutional filtering order. Traditional Convolutional Neural Networks (CNNs) tend to uniformly sample color channels, which is effective for images with rich color information, but not suitable for our data.[33, 34] To address this issue, we designed a novel image processing approach that uses longitudinal size information instead of color channel information, which can improve the correlation between channels, reduce unnecessary calculations, and improve data processing efficiency.[35, 36, 37] Therefore, the size of the preprocessed data was N × 300 × 1 × 300, representing the image number, image width, image color channel, and image length. We then compared the impact of choosing different numbers of images on model training accuracy and speed and ultimately determined that selecting 8 images was the more appropriate choice (see Supplementary

Information for the number of images selected and speed management, S3 and S4). Finally, we conducted all data preprocessing before model training to reduce the time. We have also applied data augmentation techniques to the training data, including adjusting image curves, cropping images, and scaling images, which effectively increase data diversity and improve the model's generalization ability. This process resulted in the acquisition of approximately 360,000 NumPy arrays with each sample approximately 704 KB in size.

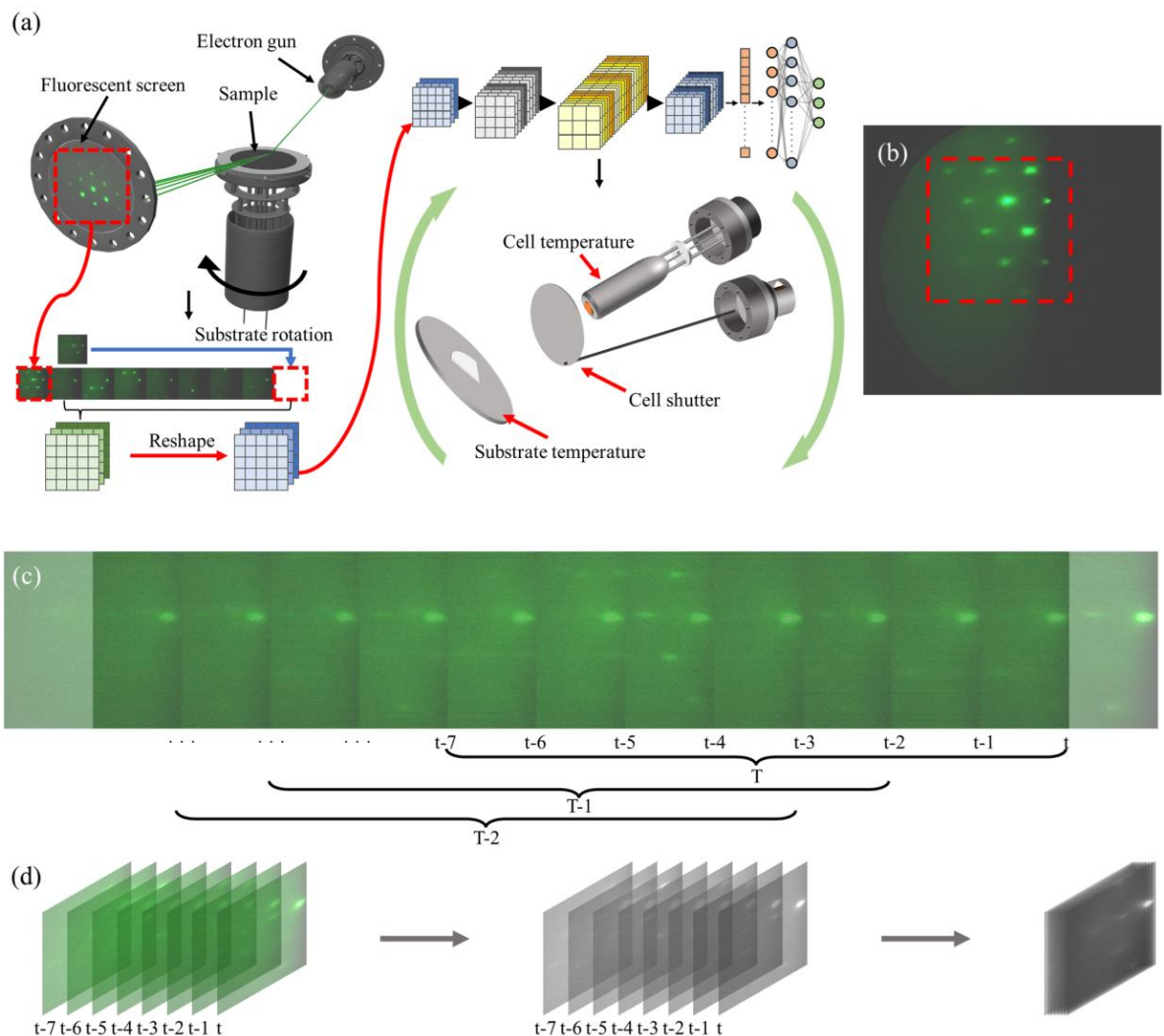

**Fig. 2: Control program framework and video processing principle.** (a) The framework of the program. (b) A typical RHEED image taken from the camera and the cropping area. (c) Continuous sampling method for RHEED images. (d) Processing method for sampled images.

**Model construction and evaluation**

The ResNet 50 model has emerged as our preferred choice for several reasons, including its advantages in the capability to automatically learn methods for extracting valuable features from raw data without the need of manual feature engineering or dimensionality reduction, rapid convergence, capability to address gradient vanishing issues, and suitability for deep training. When processing video frame data, 3D convolution proves more effective in extracting spatial and temporal dimension information from videos than the 2D convolution used in the standard ResNet model. The difference between our network and the original ResNet lies in the dimensionality of the convolutional kernel and batch normalization operations.[23] Our 3D ResNet 50 model employs 3D convolutions and 3D batch normalization. This spatiotemporal-feature-learning method enables the 3D ResNet 50 model to better capture the dynamic information in videos compared to 2D CNN, resulting in an enhanced ability to identify and differentiate the categories of videos.[38,39] Additionally, we have adjusted the structure of identity shortcuts to reduce information loss during downsampling. Furthermore, we have reduced the frequency at which the model doubles the number of channels after passing through the residual structure (see Supplementary Information for the detailed structure of the ResNet 50 model and the principle of the basic residual block in ResNet, S5 and S6). We also compared the ResNet 50 model and other models, and the results revealed that ResNet 50 outperforms the other models significantly (see Supplementary Information for comparison of training and validation results of different models, S7).[23]

We trained models for judging the QDs formation and classifying density, which are called the "QDs model" and the "density model", respectively (see Supplementary Information for model development and training result, S8). As shown in Figure 3a, the output of the QDs model is binary, represented as "yes" or "no", indicating whether the QD has formed. In contrast, the output of the density model primarily consists of "zero", "low", "medium", and "high" labels, corresponding to different QD density prediction results.

The accuracy and loss of the QDs and density models during the training are shown in Figures 3b and 3c. Overall, the improvement trend of accuracy for the two models is evident. Throughout the model training process, the persistent fluctuations and plateaus in the "QDs model" and "density model" accuracy were attributed to the data diversity and the manual label assignment process instead of limitations inherent in the model itself. After 120 training cycles, the "QDs model" maintained a validation accuracy of over 90%, as shown in Figure 3b, ultimately achieving a final average validation accuracy of 94.4%. The fluctuations observed in the "QDs model" validation process predominantly originate from a specific time before the QD formation when the RHEED streaks still dominated most of the collected images. As the QDs are about to form, the spot features closely associated with QD formation are often faint and overshadowed by the streak features. As depicted in Figure 3c, the validation accuracy of the "density model" continues to exhibit fluctuations even after 100 epochs, primarily attributed to label overlap when low-density QD data exhibit an increase in density with deposition.[40, 41] However, it consistently maintained a validation accuracy of over 80%, and the final average validation accuracy of the selected model reached 95.5%.

Furthermore, we receive multiple results from the QDs model and density mode outputs when deploying the model. We tally the results with the highest output probability to mitigate the influence of randomness and variability. This approach can reduce the likelihood of model operational errors during software deployment.[42, 43, 44] The result with the highest probability is selected to determine whether the adjustment to the growth parameters is necessary. Moreover, we converted the model into the open standard Open Neural Network Exchange (ONNX) format. When deployed in a LabVIEW environment equipped with TensorRT accelerators, the QDs model and the density model can each produce results at an approximate rate of 1 sample per second, effectively meeting the application's requirements (see Supplementary Information for hardware wiring scheme for model deployment and program interface and deployment environment, S9 and S10). Additionally, the model is universally applicable to other systems with different wobble characteristics without a new training cycle (see Supplementary Information for the QDs model and density model with different wobble characteristics, S11).

To enable early intervention in the material growth process before the QDs formation, we developed a control logic for the LabVIEW control program, as shown in Figure 3d. After opening the shutter, the substrate temperature remains unchanged if the label output from the density model aligns with the preset target before QD formation. If the output label's density from the density model exceeds the predefined target density, the substrate temperature increases; otherwise, it decreases. Once the QDs model recognizes QD formation, the shuttle will remain open until the density model's output matches the preset target. At this point, it closes the shutter to complete the growth process. This control logic adjusts the substrate temperature continuously, gradually increasing the likelihood of growing QDs with the desired target density.

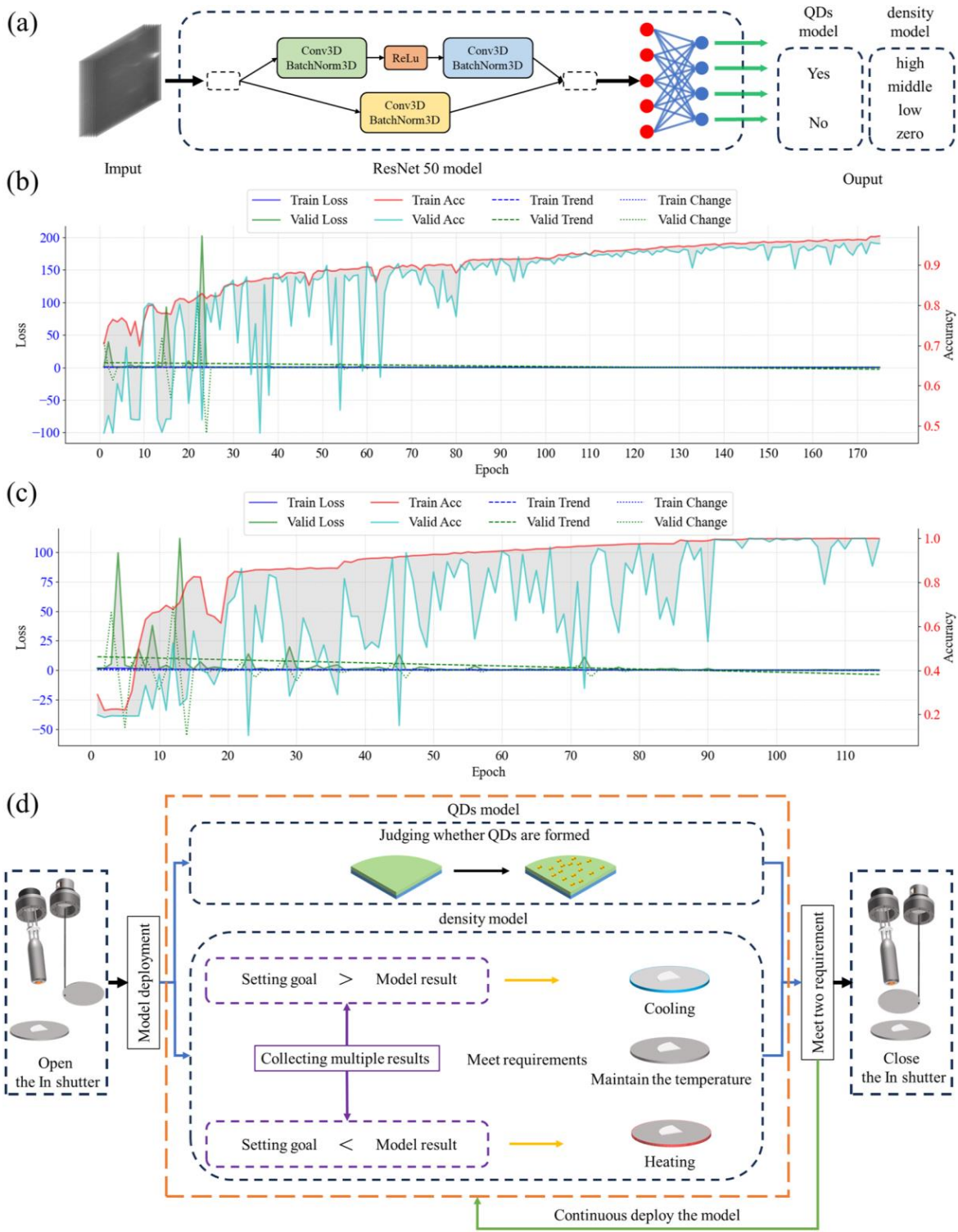

**Fig. 3:** (a) A simplified architectural diagram of the model; Training performance of (b) "QDs model", and (c) "density model": accuracy and loss as a function of trend variation (d) A control logic diagram for model deployment.

**Controlled growth of low-density QDs**

We have also grown a reference sample with QD density of $1.5 \times 10^{10}$ cm$^{-2}$, corresponding to label "middle" (see Supplementary Information for details of the reference sample, S12). Subsequently, we set the initial substrate temperature equivalent to that used in the reference sample and conducted in situ control experiment with the "low" or "high" label as the target (see Supplementary Videos for the experiments). After growth, we compiled each frame of the RHEED image captured during the growth into a sequence and analyzed. In order to distinguish different growth stages under model control, we marked the QD formation time and the In shutter closing time judged by the model in the sequence, with blue and yellow lines in the Figures 4 and 5.

As shown in Figure 4a, the substrate temperature increased by 44 °C from the beginning of the sequence until the In shutter was closed. This indicates that the initial substrate temperature was not suitable for the "low" label. In addition, the blue line and yellow line are in the same sequence, indicating that when the QDs are formed, the "density model" has already determined that the RHEED sequences are consistent with the "low" label. As shown in Figure 4b, the RHEED images mainly presents streak in the initial growth stage of QDs; it kept even before the QD formation, as shown in Figure 4c. This is attributed to the relatively flat surface of the sample in the early stage of QD growth. After the QD formation, a pattern coexisting streaks and spots was immediately observed as shown in Figure 4d, consistent with the "low" label characteristics.[27] The QDs has a density of $3.8 \times 10^8$ cm$^{-2}$, with an average diameter of 50.5 nm and height of 10.0 nm, as shown in the AFM image in Figure 4e. The result confirms the accuracy of the "density model" prediction in achieve a QD density labeled as "low".

To effectively depict changes in data trends, we have incorporated four pie charts throughout to illustrate the evolution of the model's output results, covering a statistical range spanning the initial 200 sequences, from the 200$^{th}$ sequence to 40 sequences before QD formation, the final 40 sequences before QD formation, and 40 sequences after QD formation. We classified RHEED videos as the "Yes" or "No" based on observations of QDs formation. At the beginning of growth, as shown in the first pie chart in Figure 4f, the primary output of the QDs model was "No", and the probability of outputting "Yes" was only 7%. It increased to 15% in the second pie chart although still relatively low. When the growth process approaches the formation of QDs, we can observe a significant increase from 15% to over 27% in the probability of outputting "Yes" in the third pie chart. This trend continues until the shutter closes, as shown in the fourth pie chart with over 55% in the probability of outputting "Yes". It indicates that the QDs model is susceptible to the emergence of low-density QDs (see Supplementary Information for another controlled growth experiment of low-density QDs, S13).

Figure 4g shows the "density model" results of the experiment with the "low" label set as the target. In the first pie chart, the probability of the model output "low" is significantly lower than the combined probabilities of "high" and "middle". However, with the real-time control of growth conditions, it is evident in the second pie chart that the probability of "low" output approaches 50%. The growth conditions have been gradually adjusted to better suit the growth of low-density QDs. As growth going on, it also becomes apparent from the third and fourth pie charts that the probability of the model outputting 'low' surpasses 50% and persists until the end of the growth.

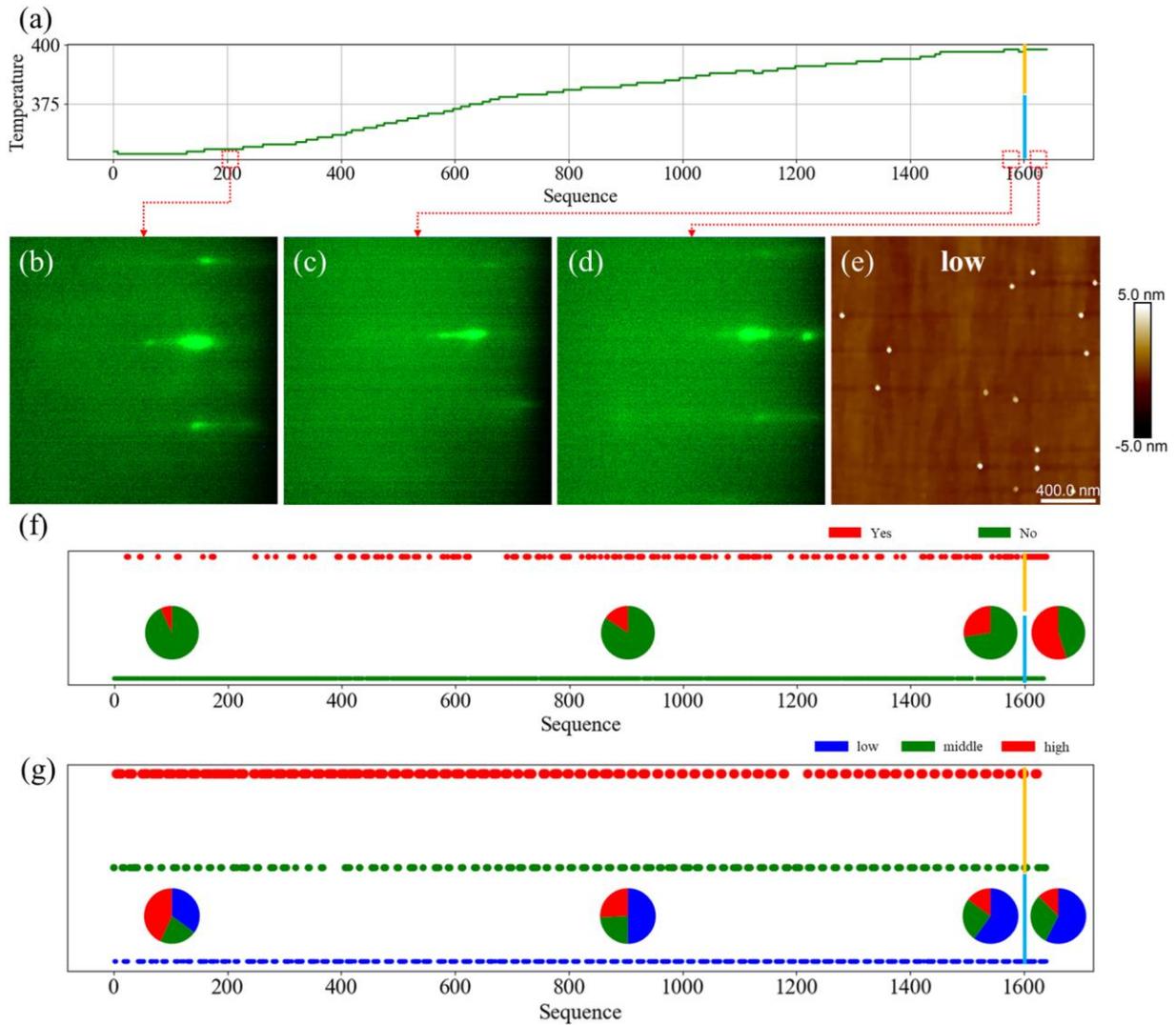

**Fig. 4: Controlled growth process of low-density QDs.** Experiment with the "low" label as the target, blue line: the QD formation time; yellow line: the In shutter closing time. (a) Substrate temperature during growth. The RHEED image (b) captured at 200$^{th}$ frame after growth; (c) before the QD formation; (d) after the QD formation. (e) The 2 μm × 2 μm AFM image of the sample. (f) The prediction results of the "QDs model"; (g) The prediction results of the "density" model.

**Controlled growth of high-density QDs**

We also conducted an in-situ control experiment with the "high" label as the target. From Figure 5a, it can be observed that the substrate temperature decreased by a total of 24 °C, indicating that the initial substrate temperature was not suitable for "high" label. In addition, the blue line in the figure did not overlap with the yellow line, which indicates the RHEED image with "high" label did not appear during the initial stage of the QD formation. In the initial stage of InAs growth, the RHEED pattern primarily showed streak patterns, as shown in Figure 5b. As the deposition amount increased, the streaks gradually became blurred before the QD formation, indicating that a certain amount of InAs has already deposited and caused the substrate surface to become rough, as shown in Figure 5c. The RHEED pattern in Figure 5d showed streaks and spots after the QD formation like those shown in Figure 4d, which are typical features of the "low" label. This explains why the blue line and the yellow line in Figure 5a did not coincide. As the In shutter closed, and the RHEED pattern exhibited spots, as shown in Figure 5e. The QDs density is approximately $1.4 \times 10^{11}$ cm$^{-2}$ with an average diameter of 33.5 nm and height of 3.2 nm, as shown in Figure 5f.

In high-density growth experiments, the shutter remains open after the formation of QDs for a period. Therefore, we incorporated two additional pie charts: one for the 40 sequences before the shutter closure and another for the 40 sequences after the shutter closure. As depicted in Figure 5g, the QDs model predominantly outputs "No" before QDs formation, with the probability of outputting "Yes" reaching a maximum of 30%, as evidenced in the first and second pie charts. As we approach the formation of QDs, the probability of the QDs model outputting "Yes" approaches 100%, as exemplified in the third through sixth pie charts in the Figure 5g, indicating high accuracy in recognizing spots pattern after the QD formation. It is noteworthy that there is still a distance between the last "No" label and the blue line, since we confirm the QD formation by collecting multiple results of the model.

The output results of the density model are also presented in Figure 5h. Before QDs formation, the density model rarely outputs "high". However, as the growth process approaches the stage of QD formation, the probability of "high" label output slowly increases to 15%, as demonstrated in the first to third pie charts. Following the formation of QDs, the probability of the density model outputting "high" rapidly increases to 30%, as depicted in the fourth pie chart. Due to the unique nature of QD systems, closing the shutter at this point would restrict the formation of high-density QDs.[45, 46] As the growth process continues and approaches the moment the shutter is closed, it becomes evident that the probability of the model outputting "high" rapidly increases to over 70%, illustrated in the fifth and sixth pie charts. This also indicates the successful growth of high-density QDs in our experiment.

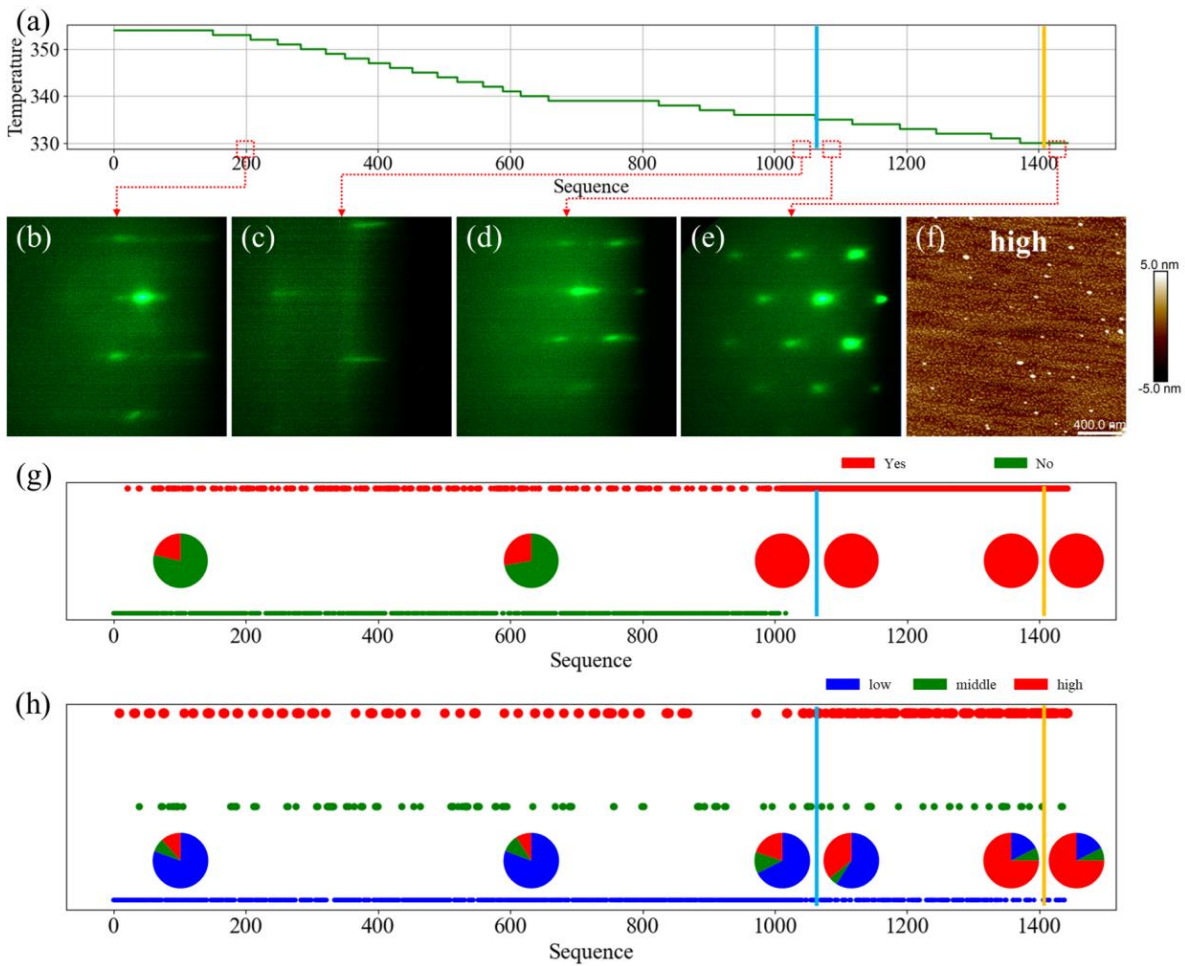

**Fig. 5: Controlled growth process of high-density QDs.** Experiment with the "high" label as the target, blue line: the QD formation time; yellow line: the In shutter closing time. (a) Substrate temperature during growth. The RHEED image (b) captured at 200$^{th}$ frame after growth; (c) before the QD formation; (d) after the QD formation; (e) before the end of the growth process. (f) The 2 μm × 2 μm AFM image of the sample. (g) The prediction results of the "QDs model", pie charts in the inset representing the probability of the results predicted; (h) The prediction results of the "density model" with the "high" label as the target.

## Discussion

Furthermore, we set the initial conditions favorable for the growth of high-density QDs and conducted experiments with the "low" label as the target (see Supplementary Information for controlled growth of low-density QDs with initial high-density growth conditions, S14). Note our experiments have a certain failure rate due to data labeling, the accuracy of the model, and limited dataset.[47, 48, 49] In our experiment, each prediction takes a few seconds, which is limited by the graphics card of our setup; the communication between the server and MBE controller takes seconds which is determined by wiring, the baud rates of serial communication, and the output power of the controller. By upgrading the hardware of our MBE system, a few tens of milliseconds of response can be achieved using the same workflow developed in this report. Moreover, as the breakthroughs in data label partitioning methods and dataset expansion in the future, we anticipate achieving finer-grained density regulation during the material growth process.

In conclusion, we developed a metrology to control the material's properties grown by MBE in situ. Specifically, we showed that a neural network could accurately predict the post-growth density of QDs within a wide range by utilizing RHEED videos information of as-grown samples.

We applied the method to InAs/GaAs QDs by tuning growth parameters in near real-time and validated it to in situ control the QD density, which was usually only possible by trial-and-error. The capability can significantly reduce the time and the number of experimental iterations required for development and optimization to achieve material specifications. In this proof-of-concept study, we trained a network to simulate a specific type of MBE growth task by varying the substrate temperature. In practice, one would train the network to the tasks with a full range of system metadata. In this way, the network would accrue an "understanding" of the complex relationships among the numerous sample and system parameters that affect the growth outcomes. Our investigations highlighted the considerable strength and practicality of the growth metrology to achieve high-quality epitaxial thin films for various applications. Moving forward, there is a huge potential for such networks to be deployed for defect detection, identification, and repair during material growth at an early stage.

## Methods

**Material growth**

The InAs QD samples were grown on GaAs substrates in the Riber 32P MBE system, equipped with arsenic (As) valved cracker, indium (In), and gallium (Ga) effusion cells. $As_4$ was used in the growth by keeping the cracker temperature at 600 °C. The beam equivalent pressure (BEP) was used to evaluate the flux of the III and V elements and calibrate their ratio.[50] A C-type thermocouple measured the substrate temperature, and the growth rates were calibrated through the RHEED oscillations of extra layers grown on GaAs substrates. The substrate was outgassed at 350 °C in the buffer chamber, then heated to 620 °C in the growth chamber for deoxidation. The substrate was then cooled to 600 °C for the growth of the GaAs buffer layer subsequently. The sample structure is shown in Figure 1a. The BEP of In was $1.4 \sim 1.5 \times 10^{-8}$ Torr, while the BEP of As was $0.95 \sim 1.0 \times 10^{-7}$ Torr. The QDs were formed in the SK growth mode after depositing 1.7 ML of InAs at a temperature of $450 \sim 500$ °C and a rate of 0.01 ML/s. The initial growth showed a relatively flat planar structure, pertaining to the wetting layer growth.[51, 52, 53, 54, 55, 56] The formation of InAs QDs was verified by the streaky-spotty transition.[57]

**Material characterization**

RHEED in the MBE growth chamber enabled us to analyze and monitor the surface of the epilayer during the growth. RHEED patterns were recorded at an electron energy of 12 kV (RHEED 12, from STAIB Instruments). A darkroom equipped with a camera was placed outside the chamber to continuously collect RHEED images with the substrate rotating at 20 rpm. The exposure time was 100 ms, with a frame sampling rate of 8 frames per second. To achieve a clear correlation of RHEED with QD density, we also characterized the surface morphology of the InAs QDs using AFM (Dimension Icon, from Burker).

**Hardware wiring scheme**

Our model is deployed on a Windows 10 computer system with an AMD R9 7950X CPU, 64GB of memory, an NVIDIA 3090 graphics card, and a 5TB solid-state drive. The program primarily obtains information from the temperature controllers, shutter controllers, and the camera to control temperature and shutter (Figure S11). We use a USB 3.0 interface to connect the camera to the computer. Temperature controllers are connected in series using the Modbus protocol to simplify the wiring. To reduce the delay in substrate temperature control, our program solely looks up information by address from the temperature controllers connected to the substrate.

**Model construction environment**

The model development occurred in the Jupyter Notebook environment on the Ubuntu system, utilizing Python version 3.9. We installed PyTorch based on CUDA 11.8 in this environment to leverage graphics card operations. The original video data was rapidly converted into the NumPy arrays format using code with random data augmentation within the Ubuntu system. Subsequently, the processed data was stored on our computer's hard disk.

**Program interface and deployment environment**

The program was developed using LabVIEW, the built-in NI VISA, NI VISION, and Python libraries for data acquisition and processing (see Supplementary Information for program interface and deployment environment, S12). The program also employs ONNX for model deployment and TensorRT for inference acceleration and allows users to set targets for the desired QD density. Once the substrate and In cell temperatures have stabilized, the program can be initiated. Firstly, the program checks the shutter status and controls the substrate and cell temperature. Once the In shutter is open, the model starts to analyze RHEED data in real-time. The model outputs are displayed as numerical values at the top of the interface and are converted to corresponding label characters on the right side. The growth status can be displayed in the "Reminder Information". Before and after the QD formation, the "Reminder Information" will show up as "Stage 1" and "Stage 2", respectively. It will display as "Finished" when the model results meet the targets several times after the QD formation.

## Data availability

The datasets generated during and/or analysed during the current study are available in the [NAME] repository, [PERSISTENT WEB LINK TO DATASETS].

## Code Availability

The code that supports the findings of this study is available from the corresponding authors upon reasonable request.

## Acknowledgements

This work was supported by the National Key R&D Program of China (Grant No. 2021YFB2206503), National Natural Science Foundation of China (Grant No. 62274159), the "Strategic Priority Research Program" of the Chinese Academy of Sciences (Grant No. XDB43010102), and CAS Project for Young Scientists in Basic Research (Grant No. YSBR-056).

## Contributions

C. S. and W. K. Z. contributed equally. C. Z. conceived of the idea, designed the investigations and the growth experiments. C. S., W. K. Z. and M. Y. L. performed the molecular beam epitaxial growth. C. S., K. Y. X., H. C., and C. X. did the sample characterization. C. S., C. Z., Z. Y. S., J. T., Z. F. W, Z. M. W., and C. L. X. wrote the manuscript. C. Z. led the molecular beam epitaxy program. B. X. and Z. G. W. supervised the team. All authors have read, contributed to, and approved the final version of the manuscript.

## Corresponding Author

*Email: zhaochao@semi.ac.cn

## Ethics declarations

### Competing interests

The authors declare no competing interests.

# Supplementary Information

Details on RHEED characteristics of QDs with different labels, the cropping area for RHEED images, the number of images selected, speed management, comparison of training and validation results of different models, the detailed structure of the ResNet 50 model, the principle of the basic residual block in ResNet, model development and result, the QDs model and density model handle data with different wobble characteristics, hardware wiring scheme for model deployment, program interface and deployment environment, details of the reference sample, Another controlled growth of low-density QDs, Controlled growth of low-density QDs with initial high-density growth conditions, comparison of QDs model and density model results with different data volumes. The video demonstrated QD growth experiments with "low" and "high" labels as targets, respectively.

Table of Contents Graphic

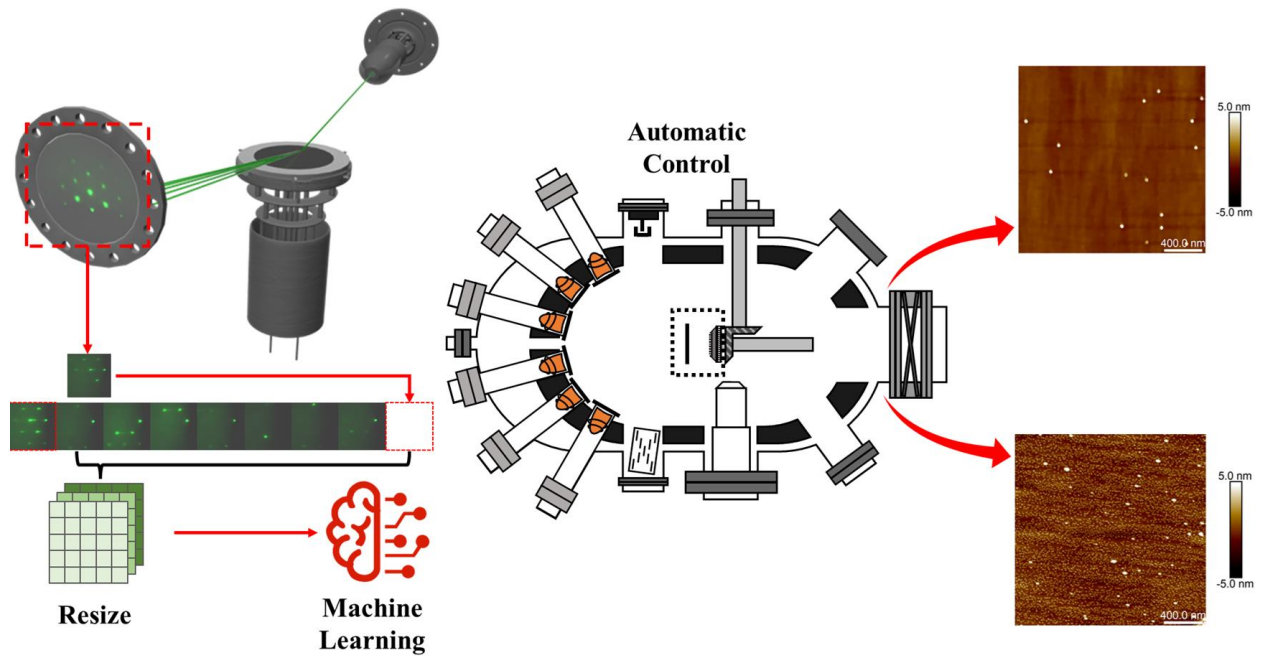